\documentclass[preprint,showpacs,preprintnumbers,amsmath,amssymb,11pt]{revtex4}

\usepackage{graphicx}
\usepackage{dcolumn}
\usepackage{bm}
\usepackage{amssymb}

\def\bea{\begin{eqnarray}}
\def\ena{\end{eqnarray}}

\begin{document}

\title{The surfing effect in the interaction of electromagnetic and gravitational waves. Limits on the speed of gravitational waves.}

\author{A. G. Polnarev}
\affiliation{ Astronomy Unit, School of Mathematical Sciences
Queen Mary, University of London, Mile End Road, London E1 4NS, UK
\footnote[1]{e-mail: A.G.Polnarev@qmul.ac.uk} }

\author{D. Baskaran}
\affiliation{School of Physics and Astronomy, Cardiff University,
Cardiff CF24 3AA, UK \footnote[2]{e-mail:
Baskaran.Deepak@astro.cf.ac.uk}}



\begin{abstract}

In the current work we investigate the propagation of
electromagnetic waves in the field of gravitational waves.
Starting with simple case of an electromagnetic wave travelling in
the field of a plane monochromatic gravitational wave we introduce
the concept of surfing effect and analyze its physical
consequences. We then generalize these results to an arbitrary
gravitational wave field. We show that, due to the transverse
nature of gravitational waves, the surfing effect leads to
significant observable consequences only if the velocity of
gravitational waves deviates from speed of light. This fact can
help to place an upper limit on the deviation of gravitational
wave velocity from speed of light. The micro-arcsecond resolution
promised by the upcoming precision interferometry experiments
allow to place stringent upper limits on $\epsilon = (v_{gw}-c)/c$
as a function of the energy density parameter for gravitational
waves $\Omega_{gw}$. For $\Omega_{gw} \approx 10^{-10}$ this limit
amounts to $\epsilon\lesssim 2\cdot 10^{-2}$.

\end{abstract}



\maketitle



\section{\label{sec:introduction} Introduction}

The detection of gravitational waves is arguable one the most
important outstanding experimental challenges in physics. With the
construction of laser interferometric gravitational wave detectors
like LIGO, VIRGO, TAMA, GEO600 there are good chances of direct
detection in the very near future \cite{Thorne1987},
\cite{Allen1997}, \cite{glpps2001}, \cite{CutlerThorne2001},
\cite{Hughes2003}, \cite{Grishchuk2003}, \cite{sathya2005}. Along
side the interferometers, which are mainly aimed at detecting
gravitational waves of astrophysical origin, the anisotropies in
temperature and polarization of the Cosmic Microwave Background
(CMB) have a strong potential to discover relic gravitational
waves \cite{grishchuk1974}, \cite{Basko1980}, \cite{Polnarev1985}
(see \cite{Keating2006a}, \cite{dgp2006} for a recent reviews).

Most of the current techniques to detect gravitational waves are
based on their interaction with electromagnetic fields. In
general, the interaction of gravitational waves with
electromagnetic radiation leaves imprints on the latter that can
be experimentally measured \cite{GrishchukPolnarev1980}. In this
work we shall deal with one such interaction effect which we shall
call the ``surfing effect", where (figuratively speaking) the
electromagnetic wave surfs on a gravitational wave leading to an
observable phase change in the electromagnetic wave. This effect
was first considered in \cite{bkpn1990}. In the present paper, we
shall expand on the results of \cite{bkpn1990}, \cite{bkpn1992}
and generalize the effect for an arbitrary gravitational wave
field. We shall consider the consequences of this effect for the
planned precision radio (or x-ray) interferometric projects
\cite{RadioAstron}, \cite{Interferometers}. As we shall show, due
to the transverse nature of gravitational waves, the surfing
effect leads to an observable phase change only when the velocity
of gravitational waves is different from speed of light. Using
this fact, and the micro-arcsecond accuracy promised by the
precision interferometry measurements \cite{RadioAstron},
\cite{Interferometers}, we can place significant upper limits on
the parameter $\epsilon = (c-v_{gw})/c$ which characterizes the
deviation of velocity of gravitational waves from speed of light.

The constraints on the speed of gravitational waves is an
interesting experimental challenge. A potentially strong method of
constraining $\epsilon$ is to compare the arrival times of a
gravitational wave and an electromagnetic wave emitted by a
supernova or a gamma ray burst \cite{Will1998} (see also
\cite{Will2001}). Although this method will be able to give very
strong constraints, it crucially depends on our ability to model
and detect the gravitational wave signal from these sources, which
is a significant theoretical and experimental challenge. In
\cite{Kopeikin2004} the speed of gravity, and correspondingly
indirectly the speed of gravitational waves, was constrained, by
analysis of retarded gravitational potentials in the non-wave
zone, by measuring propagation of the quasar's radio signal past
Jupiter. Let us note, without going into detail of ambiguity in
the interpretation of this result (see for example
\cite{Will2003}), that speed of gravity was constrained to $c_g =
(1.06\pm 0.21)c$ corresponding to $\epsilon \lesssim 0.2$. In this
paper we consider the propagation of electromagnetic radiation in
the field of gravitational waves, i.e.~in the wave zone of the
gravitational field, where the velocity parameter $\epsilon$ can
be introduced avoiding any ambiguity. Using the surfing effect, we
consider an independent method of placing upper limits on the
$\epsilon$-parameter, which could potentially give very strong
limits on the speed of gravitational waves ($\epsilon\lesssim
10^{-2}$ for $\Omega_{gw}\approx 10^{-10}$).

%

The plan of the paper is as follows. In Section
\ref{sec:singlewave} we shall consider the propagation of an
electromagnetic wave in the field of a single monochromatic
gravitational wave. We shall discuss the physical aspects of the
surfing effect with a view on the precision interferometric
measurements, as well as write down some of the equations that
will be used in the following sections. In Section
\ref{sec:arbitrarywavefield} we generalize the surfing effect for
an arbitrary gravitational wave field. Positing statistical
properties of the gravitational wave field we derive the
consequent statistical properties of the response of an
interferometer. In Section \ref{sec:upperlimit} we use the surfing
effect along with the predicted precision level of the
interferometry measurements to place upper limits on the velocity
parameter $\epsilon$ depending on the energy density of
gravitational wave described by the density parameter
$\Omega_{gw}$. Finally, in Section \ref{sec:conclusion} we present
a short discussion and summary of the main results of this paper.

\section{\label{sec:singlewave} Single monochromatic gravitational wave}

Let us consider a slightly perturbed flat
Friedmann-Lema\^itre-Robertson-Walker (FLRW) universe with
coordinates $x^{\mu}\equiv (\eta,x^i)$ and the metric given by
\cite{LandauLifshitz}, \cite{mtw}
\bea
ds^2 = g_{\mu\nu}dx^{\mu}dx^{\nu} = a^2(\eta)\left[-d\eta^2 +
\left(\delta_{ij} + h_{ij}\right)dx^idx^j\right].
\label{metric}
\ena
Here $\eta$ is conformal time coordinate related to proper time
coordinate $t$ through a relation $cdt = a(\eta)d\eta$, and $x^i$
are the spatial coordinates. In the above expression $a(\eta)$ is
the scale factor, and $h_{ij}$ is the gravitational wave
perturbation. Since, in this work we shall be interested in
electromagnetic waves emitted by objects with a redshift
$z\lesssim5$, we will work with a matter dominated Universe model
characterized by a scale factor
\bea
a(\eta) = a_o\left(\frac{\eta}{\eta_o}\right)^2,
\label{mattermetric}
\ena
where $\eta_o$ corresponds to present time. From the above
expression it follows that $a_o$ is the scale factor at the
present time. Without loss of generality we can set $a_o = 1$.
With this convention, the value of $\eta_o$ is related with the
present day Hubble constant $H_o$
\bea
\eta_o = \frac{2c}{H_o} = 2L_H,
\ena
where we have introduced the Hubble length $L_H \equiv c/H_o$.

For simplicity, in this section we shall consider the case of a
single monochromatic plane gravitational wave. We shall also
restrict our considerations to gravitational waves whose
wavelength at the current epoch is small compared to the present
Hubble length. This implies that we can consider these waves to be
monochromatic waves with an amplitude damping adiabatically with
the expansion of the universe
\bea
h_{ik}(\eta,x^i) = h_o\left(\frac{a_o}{a(\eta)}\right)~ p_{ik}~
e^{ik_{\mu} x^{\mu}},\label{planegravwave}
\ena
where $h_o$ is the amplitude at the present epoch, $p_{ik}$ is the
polarization tensor of the gravitational wave, and $k_\mu =
\left(k_0,k_i\right)$ is the wave vector of the gravitational wave
\cite{LandauLifshitz}. It is convenient to introduce the
wavenumber $k = \left(\delta_{ij}k^ik^j\right)^{1/2}$. The present
day wavelength of the gravitational field $\lambda_{gw}$, is
related to wavenumber $k$ by the relation $\lambda_{gw} = 2\pi/k$.
The present day frequency of the gravitational wave $\omega_{gw}$
is related to the time component of the wavevector $k_0$ by
relation $\omega_{gw} = ck_0$.

Let us analyze the electromagnetic wave propagation in the
approximation of geometrical optics. In this approximation, the
wave equation is written for the quantity $ \psi(x^{\mu}) \equiv
\psi (\eta,x^i)$, known as the eikonal \cite{LandauLifshitz}.
Eikonal has the physical meaning of the phase of an
electromagnetic wave field $ f(x^{\mu}) =
A(x^{\mu})e^{i\psi(x^{\mu})}$. (The quantity $A(x^{\mu})$
describes the amplitude of the wave field, but shall not interest
us in this consideration.) The wave vector $\kappa_{\mu}$ of the
electromagnetic wave is give by $\kappa_{\mu} = \frac{\partial
\psi}{\partial x^{\mu}}$.

The eikonal equation follows from the isotropy condition
$g_{\mu\nu}\kappa^{\mu}\kappa^{\nu} = 0$. Substituting the
expression for $\kappa^{\mu}$ in terms of $\psi$ we arrive at the
eikonal equation \cite{LandauLifshitz}
\bea
g^{\mu\nu} \frac{\partial \psi}{\partial x^{\mu}}\frac{\partial
\psi}{\partial x^{\nu}} = 0. \label{eikonalequation}
\ena
We shall seek the solution to equation (\ref{eikonalequation}) in
a perturbative form
\bea
\psi(x^{\mu}) = \psi_0(x^\mu) + \psi_1(x^{\mu}).
\label{perturbativesolution}
\ena
Here the zeroth order solution corresponds to the solution in the
absence of perturbations, while the first order solution
corresponds to the solution in the presence of perturbations $\sim
h$. Let us assume that the zeroth order solution corresponds to a
plane monochromatic electromagnetic wave:
\bea
\begin{array}{c} \psi_0 = {\omega_E} \left(\eta +  e_ix^i\right)/{c},\\ \frac{\partial
\psi_0}{\partial \eta} = {\omega_E}/{c}, ~~~\frac{\partial
\psi_0}{\partial x^i} = {\omega_{E} e_i}/{c}, \end{array}
\label{zerothordersolution}
\ena
where $e^i$ ($\delta^{ij}e_ie_j=1$) is the unit vector along the
direction of wave propagation, and $\omega_{E}$ is the frequency
of the electromagnetic wave at the present time. Taking into
account the solution (\ref{zerothordersolution}) of the zeroth
order equation, the first order equation for $\psi_1(x^{\mu})$
takes the form
\bea
\frac{\partial \psi_1}{\partial \eta} + e^i\frac{\partial
\psi_1}{\partial x^i} =
\frac{1}{2c}\omega_{E}~h_{ik}~e^ie^k.\label{firstorderequation}
\ena
The solution to equation (\ref{firstorderequation}) is written in
terms of the line of sight integral (along the unperturbed light
path):
\bea
\psi_1(\eta,x^i) = \frac{1}{2c}\omega_{E} ~ e^ie^k
\int\limits_{\eta_o-D}^{\eta_o} ds ~
h_{ik}\left(s,x^i+e^i\left(s-\eta_o\right)\right),\label{firstordersolution1}
\ena
where $s$ is the $\eta$-time parameter along the light ray path
from the emitter to observer. The limits of integration
$\left(\eta_o-D\right)$ and $\eta_o$ correspond to the time of
emission and observation correspondingly. Using
(\ref{mattermetric}), the parameter $D$ can be related to the
redshift of emission $z$ by
\bea
D = 2L_H\left(1-\frac{1}{\sqrt{1+z}}\right).
\label{Dofz}
\ena

Below, for simplicity of analysis and in order not to obscure the
physical interpretation of the surfing effect, we shall consider
the problem in flat space-time without cosmological evolution of
the scale factor, and correspondingly, assume no cosmological
evolution of gravitational wave amplitude. This analysis is
equivalent to setting the scale factor in expression
(\ref{metric}) to a constant value, i.e.~$a(\eta) = a_o = 1$.
Alternatively, this approximation can be viewed as an analysis
restricted to small values of redshift, i.e.~$z\ll1$. In this
limit, parameter $D$ represents the physical distance to the
source, and is related to the redshift by the usual Hubble law
$D\approx zL_H$. We shall reintroduce the cosmological evolution
in Section \ref{sec:arbitrarywavefield}, where we shall explain
how and why the result modifies. Some of the calculational
subtleties that arise when analyzing this situation are considered
in Appendix \ref{CosmologyInclusion}.

Thus, assuming a plane gravitational wave with constant amplitude
(i.e.~setting $a(\eta) = a_o$ in (\ref{planegravwave})), we can
explicitly evaluate the  integral in (\ref{firstordersolution1})
to get the resulting phase change due to a gravitational wave
\bea
\psi_1(\eta,x^i) = \frac{\omega_E  h_o}{2c} ~
e^ie^kp_{ik}~e^{-i\left(k_0\eta-k_ix^i\right)}
~\left(\frac{e^{iD\left(k_0-k_ie^i\right)}-1}{i\left(k_0-k_ie^i\right)}\right).\label{firstordersolution2}
\ena

An interferometer is an experimental device capable of measuring
the difference in phase in an electromagnetic wave
\cite{Interferometers}. The micro-arcsecond precision promised by
the upcoming interferometry projects may allow to detect signature
of gravitational waves or set upper limits on their magnitude. For
this reason, let us switch our attention to the interferometers.
The interferometers measure the variation in the phase of the
electromagnetic signal received by its base antennae. Let us
consider a long base interferometer setup. In this case there are
two antennae separated by a spatial vector $L^i$. Without loosing
generality, we can assume that the first of this antennae is
located at the coordinate origin. Then, the difference of phase
measured by the by the two antennae, due to the gravitational wave
influence, is given by
\bea
\Delta\psi_1 = \psi_1(\eta,0) - \psi_1(\eta,L^i).
\label{deltapsidefinition}
\ena
For the plane electromagnetic wave under consideration,
substituting (\ref{firstordersolution2}) into
(\ref{deltapsidefinition}) we arrive at the expression for phase
difference measured by the interferometer
\bea
\Delta\psi_1 = \frac{\omega_E  h_o}{2c} ~ e^ie^kp_{ik} e^{-ik_0
\eta}
\left[\frac{e^{iD\left(k_0-k_ie^i\right)}-1}{i\left(k_0-k_ie^i\right)}\right]
\left( 1 - e^{ik_iL^i} \right)\ \label{deltapsi1}.
\ena
The output of an interferometric measurements is usually quoted in
terms of angular resolution. We shall use this convention. The
angular resolution of an interferometer corresponding to a phase
resolution $\Delta\psi_1$ is give by
\bea
\Delta\alpha = \frac{c}{L\omega_{E}}\Delta\psi_1,
\ena
where $L = \left(\delta_{ij}L^iL^j\right)^{1/2}$ is the baseline
length of the interferometer.

In what follows, we shall be interested in gravitational waves
with wavelengths considerably larger that the baseline of the
interferometer, i.e.~$L/\lambda_{gw} \ll 1$. In the case of a
space-borne interferometer with base length of $10^6~km$, this
implies $\nu_{gw}\ll 10^{-1}~Hz$ for the gravitational wave
frequency. For interferometers with a shorter baseline this limit
can be significantly larger.

To proceed, let us further introduce unit vectors $l^i = L^i/L$
and $\tilde{k}^i = k^i/k$. $l^i$ is a unit vector pointing in the
direction of the interferometer baseline, while $\tilde{k}^i$ is
the unit vector pointing in the direction of the gravitational
wave propagation. Assuming $L/\lambda_{gw} \ll 1$ in equation
(\ref{deltapsi1}), we can expand this expression in a series
retaining only the lowest order term in $k_iL^i$. We get
\bea
\Delta\alpha = \frac{1}{2}h_o~ k_rl^r ~  e^ie^kp_{ik} ~ e^{-ik_0
\eta} \left[
\frac{1-e^{iD\left(k_0-k_ie^i\right)}}{\left(k_0-k_ie^i\right)}
\right].\label{deltaalpha1}
\ena

Up to now we have not posited any relationship between the
gravitational wave frequency $ck_0$ and the magnitude of the
wavenumber $k$. This relationship (dispersion relation) defines
the velocity of a gravitational wave $v_{gw} = ck_0/k$. In General
Relativity this velocity equals the speed of light, but in an
alternative theory this might not be the case. In order to analyze
the possibility that $v_{gw}\neq c$, let us use the
phenomenological parameter $\epsilon$, introduced in the previous
section
\bea
\epsilon \equiv \frac{c-v_{gw}}{c},~~~\textrm{where}~~~v_{gw}
\equiv \frac{ck_0}{k},
\ena
which characterizes the relative deviation of velocity of
gravitational waves from the speed of light. Let us note that,
$\epsilon$ can be related to a non vanishing rest mass of a
graviton $m_g$ through the relation
\bea
\epsilon = 1 - \frac{\hbar \omega_{gw}}{\hbar
\omega_{gw}+m_gc^2}\approx \frac{m_gc^2}{\hbar \omega_{gw}}.
\label{gravitonmass}
\ena

Returning to equation (\ref{deltaalpha1}), and substituting the
relationship $k_0 = (1-\epsilon)k$ into it, we get
\bea
\Delta\alpha = \frac{1}{2}h_o ~e^ie^kp_{ik} ~ \tilde{k}_rl^r ~
e^{-ik(1-\epsilon) \eta} \left[
\frac{1-e^{ikD\left(1-\epsilon-\tilde{k}_ie^i\right)}}{\left(1-\epsilon-\tilde{k}_ie^i\right)}
\right].\label{deltaalpha2}
\ena
It is instructive at this point to look more closely at expression
(\ref{deltaalpha2}). We are considering an electromagnetic wave
travelling ``along" a plane gravitational wave. As follows from
this expression, the angular displacement $|\Delta\alpha|$ is most
pronounced when both the waves move in almost parallel directions.
This picture is reminiscent of wave surfing, and hence we call
this the ``surfing" effect. The expression in square brackets
becomes large (proportional to $kD\sim D/\lambda_{gw}$), when the
denominator tends to zero (i.e.~$1-\epsilon-\tilde{k}_ie^i
\rightarrow 0$), leading to a resonance effect. In the case when
$\epsilon = 0$, this does not lead to significant growth of the
angular displacement $|\Delta\alpha|$, because of the transverse
nature of gravitational waves (since $e^ie^jp_{ij} \rightarrow 0$
as $\tilde{k}_ie^i \rightarrow 1$). On the other hand, if
$\epsilon\neq 0$, the expression for $|\Delta\alpha|$ becomes
sufficiently large for a gravitational wave propagating at an
angle $\cos{\theta} = \tilde{k}_ie^i \approx (1-\epsilon)$ to the
line of sight. Thus, due to the transverse nature of the
gravitational waves this surfing effect is absent if $\epsilon =
0$, but can become significant for $\epsilon \neq 0$. This effect,
as we shall show in the following section, can be used to put
stringent constraints on the $\epsilon$ parameter characterizing
the velocity of gravitational waves.

\section{\label{sec:arbitrarywavefield} Arbitrary gravitational wave field}

In the previous section we considered the case of a single
monochromatic gravitational wave. In order to generalize the
considerations of the previous section, let us now consider an
arbitrary gravitational wave field. This field can be decomposed
into spatial Fourier modes
\bea
h_{ij}(\eta,x^i) = \int d^3{\bf{k}} \sum_{s=1,2} \left[
h_s(k^i,\eta)\stackrel{s}{p}_{ij}(k^l)e^{ik_ix^i} +
h_s^*(k^i,\eta)\stackrel{s}{p}_{ij}^*(k^l)e^{-ik_ix^i}
\right],\label{fouriergw}
\ena
where $d^3{\bf{k}}$ denotes the integration over all possible wave
vectors, and $s=1,2$ correspond to the two linearly independent
polarizations of a gravitational wave. The mode functions
$h_s(k^i,\eta)$ have the following time evolution
\bea
h_s(k^i,\eta) =
h_s(k^i)\left(\frac{a_o}{a(\eta)}\right)e^{-ik(1-\epsilon)
\eta},\label{hmodefunctions}
\ena
where $h_s(k^i)$ is the gravitational wave amplitude at the
present time.

Due to the linear nature of the problem, following the
decomposition (\ref{fouriergw}), the total angular displacement
due to gravitational waves can also be decomposed into Fourier
modes in a similar fashion
\bea
\Delta\alpha = \int d^3{\bf{k}} \sum_{s=1,2} \left[
h_s(k^i)\Delta\tilde{\alpha}_s(k^i) +
h_s^*(k^i)\Delta\tilde{\alpha}_s^*(k^i)
\right],\label{fourieralpha}
\ena
where we have introduced a tilde over $\alpha$, in the right hand
side of the above expression, to indicate the explicit factoring
out of the gravitational wave amplitude $h_s(k^i)$ compared with
expression (\ref{deltaalpha2}).

Using the results of previous subsection (see Eq.\
(\ref{deltaalpha2})), ignoring cosmological evolution for now, the
contribution from a single Fourier component
$\tilde{\alpha}_s(k^i)$ is given by
\bea
\Delta\tilde{\alpha}_s(k^i) = \frac{1}{2} \tilde{k}_rl^r ~
e^ie^k\stackrel{s}{p}_{ik} ~ e^{-ik(1-\epsilon) \eta} \left[
\frac{1-e^{ikD\left(1-\epsilon-\tilde{k}_ie^i\right)}}{\left(1-\epsilon-\tilde{k}_ie^i\right)}
\right].\label{fourieralphacomponentvalue}
\ena

In general for an arbitrary gravitational wave field
(\ref{fouriergw}) the angular displacement measured by the
interferometer is given by expressions (\ref{fourieralpha}) and
(\ref{fourieralphacomponentvalue}). In practice we do not have the
precise information about the gravitational wave field, and are
restricted to knowledge of only its statistical properties. Let us
assume the following statistical properties for the mode functions
$h_s(k^i)$
\bea
<h_s(k^i)> = 0,~~~ <h_s(k^i)~ h_{s'}^{*}(k'^i)> =
\frac{P_h(k)}{16\pi k^3}\delta_{ss'}\delta^3(k^i-k'^i),
\label{gwstatprop}
\ena
where the brackets denote ensemble averaging, and $P_h(k)$ is the
metric power spectrum per logarithmic interval of $k$. These
conditions correspond to a stationary statistically homogeneous
and isotropic gravitational wave field.

The statistical properties of $\Delta\alpha$ follow from the
statistical properties of the underlying gravitational wave field
(\ref{gwstatprop}). Using (\ref{fourieralpha}),
(\ref{fourieralphacomponentvalue}) and (\ref{gwstatprop}) after
straight forward calculations we get following statistical
properties for angular displacement $\Delta\alpha$:
\begin{subequations}
\bea
<\Delta\alpha> &=& 0,\label{deltaalphasquare_0}\\ <\Delta\alpha^2>
&=& \int \frac{dk}{k} P_h(k) \Delta\tilde{\alpha}^2(k),
\label{deltaalphasquare}
\ena
\end{subequations} where we have introduced the transfer function
\bea
\Delta\tilde{\alpha}^2(k) = \frac{1}{8\pi}\int d\Omega \sum_s
\left| \Delta\tilde{\alpha}_s(k^i) \right|^2.
\label{deltaalphasquare2}
\ena
In the above expression $d\Omega$ represents integration over the
possible directions of g.w. wave (i.e.~$d^3{\bf k} =
k^2dkd\Omega$).

\begin{figure}
\begin{center}
\includegraphics[width=7cm]{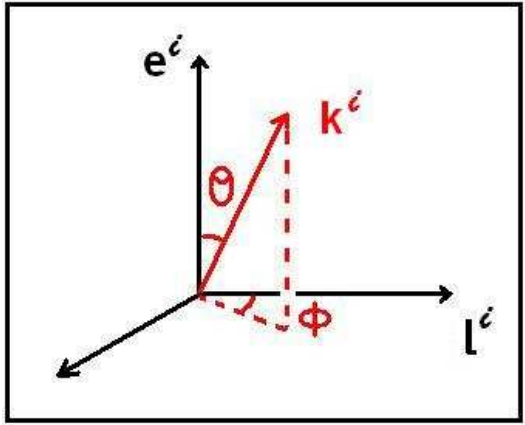}
\end{center}
\caption{The graphical representation of the various vectors and
angles used in the text. Vector $e^i$ is the unit vector along the
direction of the electromagnetic wave, unit vector $l^i$ is
aligned with the base of the interferometer, and $k^{i}$ is the
wave vector of the gravitational wave.}\label{figurecoordinates}
\end{figure}

\begin{figure}
\begin{center}
\includegraphics[width=7cm]{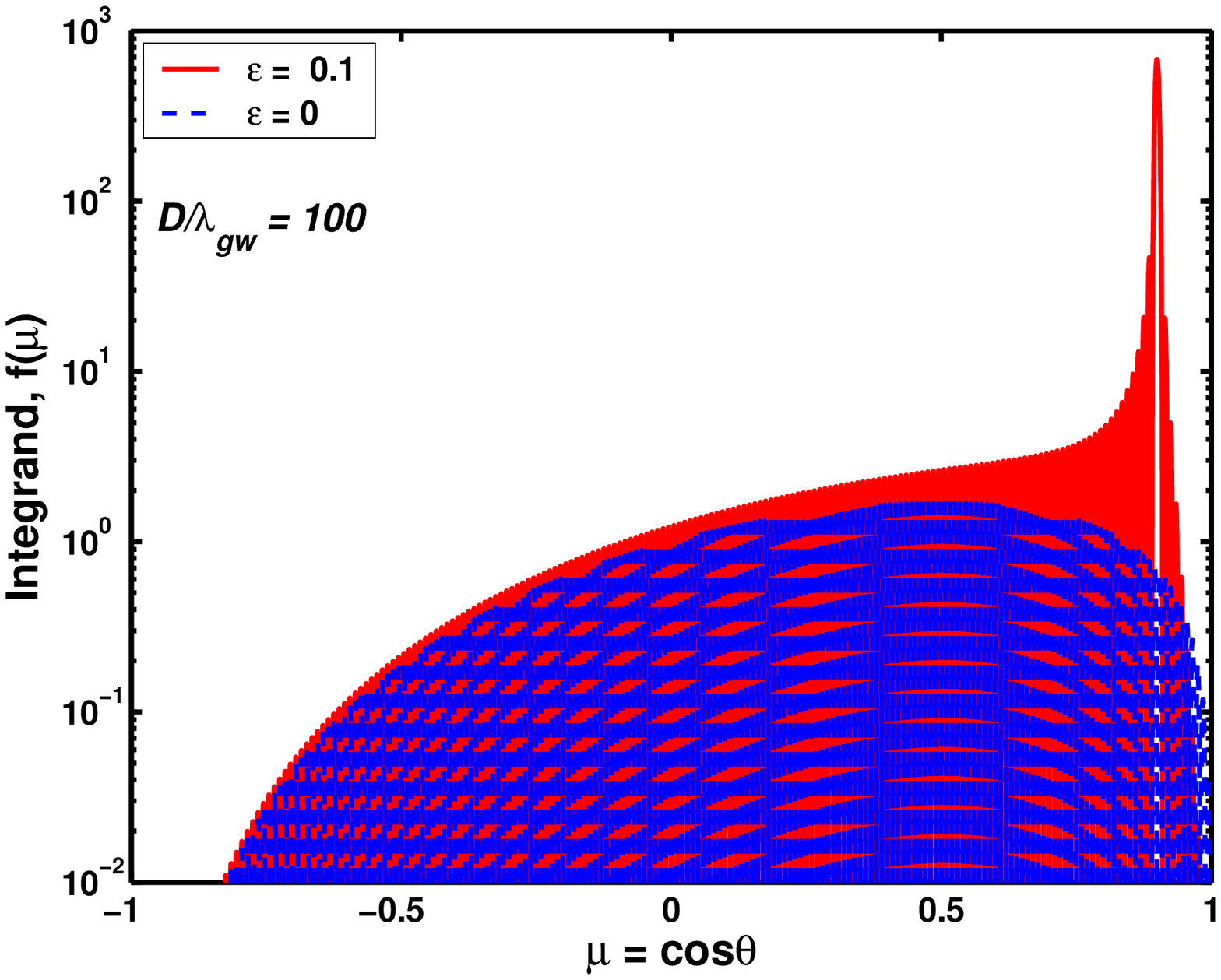}
\includegraphics[width=7cm]{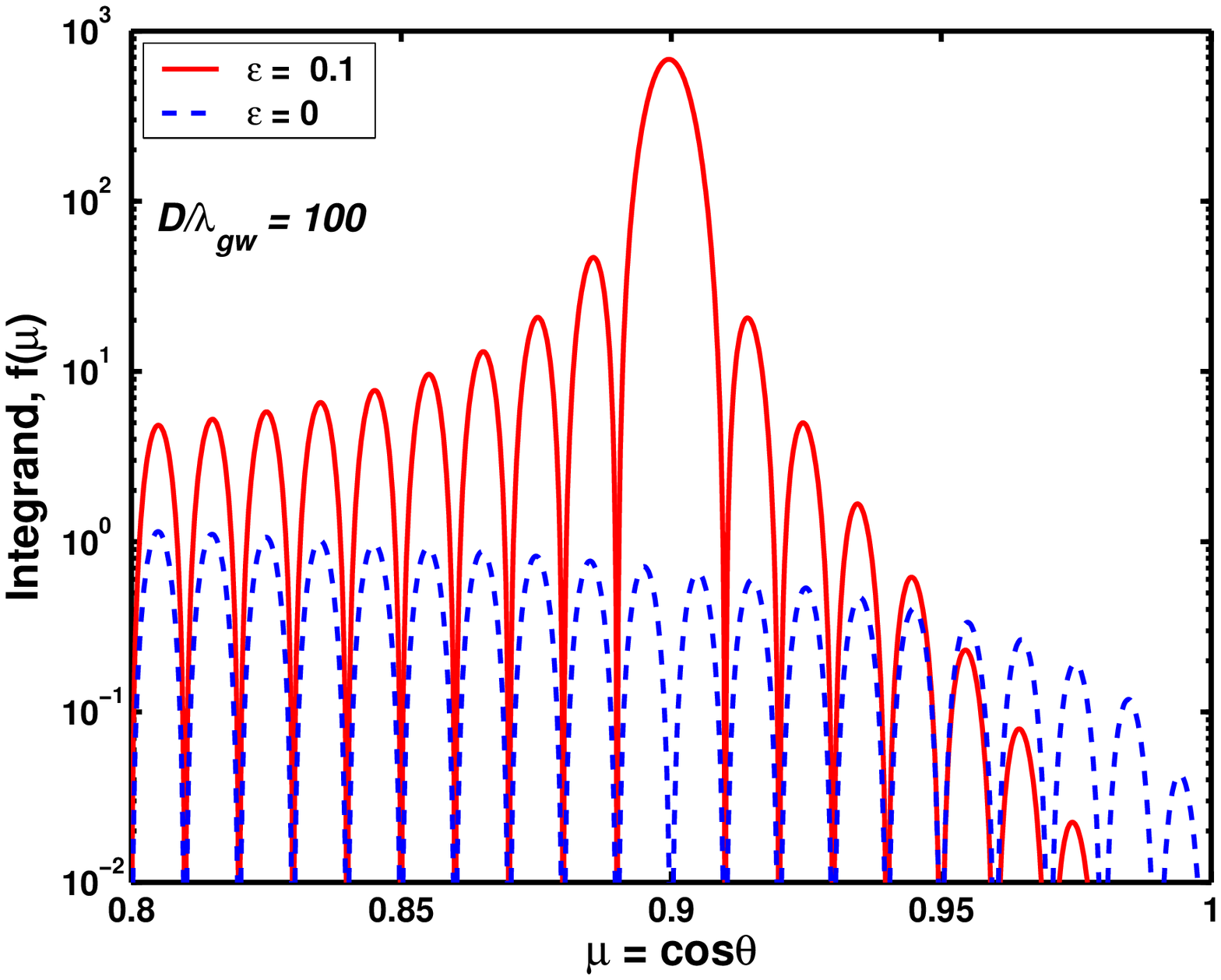}
\end{center}
\caption{The illustration of the resonance effect, present for
$\epsilon \neq 0$. The graphs show integrand in expression
(\ref{deltaalphasquare3}). For the case $\epsilon\neq 0$ the
integrand sharply peaks at angle $\mu\approx (1-\epsilon)$ (solid
red line), while for the case $\epsilon = 0$ the effect is absent
(dashed blue line). In the case of $\epsilon \neq 0$, the
gravitational waves travelling at an angle $\cos{\theta} = \approx
(1-\epsilon)$ to the line of sight are the predominant
contributors to the surfing effect. The figure on the left shows
the integrand for the whole region of $\mu$, while the figure on
the right zooms into the region around the
resonance.}\label{figure1}
\end{figure}

We shall now proceed to calculate the expression
(\ref{deltaalphasquare2}) explicitly. Let us introduce a spherical
coordinate system $(\theta,\phi)$ related to the spatial
coordinates $\{x^i\}$ in the usual manner \cite{Goldstein}.
Without loss of generality, we can assume that we are looking in
the north-pole direction, i.e.~$e^i = (0,0,1)$. Let us also
introduce the quantity $\mu = \cos{\theta} = e_i\tilde{k}^i$,
characterizing the angle between a gravitational wave and the
direction of observation. Furthermore for an interferometer, for
optimal resolution $l^i$ is aligned perpendicular to $e^i$, thus
without loss of generality we can assume $l^i = (1,0,0)$. The
geometry of the problem is presented in Figure
\ref{figurecoordinates}. The polarization tensors for
gravitational waves have the form $\stackrel{s}{p}_{ij}(k^i) =
(e^{\theta}_i\pm i e^{\phi}_i)(e^{\theta}_j\pm i e^{\phi}_j)/2$,
with $\pm$ corresponding to the two independent circularly
polarized degrees of freedom $s=1,2$ ($e^{\theta}_i$ and
$e^{\phi}_i$ are the meridian and azimuthal unit vectors
perpendicular to the gravitational wave wavevector $k_i$, for a
detailed discussion see for example \cite{dgp2006},
\cite{Baskaran2004}). Taking into account the relations
\bea
e^ie^j\stackrel{s}{p}_{ij} = \frac{1}{2}(1-\mu^2)e^{\pm
2i\phi},~~~ l_i\tilde{k}^i = (1-\mu^2)^{1/2}\cos{\phi}, \nonumber
\ena
and substituting (\ref{fourieralphacomponentvalue}) into
(\ref{deltaalphasquare2}), after straight forward manipulations,
the expression (\ref{deltaalphasquare2}) for the transfer function
takes the form:
\bea
\Delta\tilde{\alpha}^2(k) = \frac{1}{16}\int\limits_{-1}^{+1} d\mu
~ \left( 1-\mu^2 \right)^3 \left[\frac{\sin^2{\left\{ \frac{\pi
D}{\lambda_{gw}} \left(1 - \epsilon - \mu\right)
\right\}}}{\left(1 - \epsilon -
\mu\right)^2}\right].\label{deltaalphasquare3}
\ena
The terms in the above integral have a clear physical meaning. The
factor $(1-\mu^2)^3$ is due to the transverse nature of the
gravitational waves and the geometry of space interferometry
($(1-\mu^2)^2$ and $(1-\mu^2)$ terms correspondingly). The
quantity in square brackets sharply peaks at values $\mu \approx
(1-\epsilon)$, which is the result of a resonance effect,
i.e.~what we call the surfing effect, for gravitational waves
travelling at an angle $\cos{\theta}\approx (1-\epsilon)$ to the
line of sight. Due to the pre-factor $(1-\mu^2)^3$, this resonance
does not give a significant contribution for the case $\epsilon =
0$. The integrand in expression (\ref{deltaalphasquare3}) is
plotted for the two cases in Figure \ref{figure1}. In the limit
$\epsilon \rightarrow 0$ and $D/\lambda_{gw} \rightarrow \infty$
we can calculate the transfer function (\ref{deltaalphasquare3})
explicitly. Referring the reader to appendix
\ref{SurfingIntegrals} for details of calculation, let us present
the final result below:
\bea
\Delta\tilde{\alpha}^2(k) \approx \frac{1}{20}\left[1 +\frac{}{}
5\pi\epsilon^3kD\right].
\label{deltaalphasquare4}
\ena

As was mentioned previously, when deriving expression
(\ref{deltaalphasquare4}) we had ignored the cosmological
evolution of the gravitational wave amplitude. In Appendix
\ref{CosmologyInclusion} we derive the expression for the transfer
function when the cosmological evolution of gravitational waves is
properly taken into account. The resulting expression is as
follows:
\bea
\Delta\tilde{\alpha}^2(k) &\approx& \frac{1}{20}\left[
\left(\frac{(1+z)^2+1}{2}\right) +
5\pi\epsilon^3kD\left(1+z\right)\right] \nonumber \\
&=&\frac{1}{20}\left[ \left(\frac{(1+z)^2+1}{2}\right) +
5\pi\epsilon^3kL_H\left(1+z\right)\left(2-\frac{2}{\sqrt{1+z}}\right)\right].
\label{deltaalphasquare4cosmological}
\ena
From expression (\ref{deltaalphasquare4cosmological}) we can
quantify the condition for the resonance to occur by comparing the
two terms in the square brackets. The resonance occurs when, in
the right side of (\ref{deltaalphasquare4cosmological}), the
second term (resonance term) is larger that the first term
(non-resonance term), i.e.~when $\epsilon^3kL_H\gg1$. As we shall
show in the next section, given the planned level of sensitivity
for interferometric measurements, the resonance effect allows to
place significant upper bounds on the parameter $\epsilon$.

The redshift factors $z$, occurring in expression
(\ref{deltaalphasquare4cosmological}), have a clear physical
interpretation. The factor $(1+z)^2$ in the non-resonance part of
the transfer function occurs because, its main contribution comes
from epoch when the electromagnetic radiation was emitted,
corresponding to a redshift of $z$. It is thus sensitive to the
gravitational wave power spectrum at the epoch of emission, which
was a factor $(1+z)^2$ stronger than today. On the other hand, in
the case of the resonance term, the contribution to the transfer
function is gained along the path from emitter to observer. This
leads to a weaker $z$ dependence in this term compared with the
non-resonance term.

\section{\label{sec:upperlimit} Upper limit on the velocity of gravitational waves}

Let us now consider the implications of the surfing effect for the
precision interferometry measurements, and the achievable upper
limits on $\epsilon$. When considering stochastic gravitational
wave fields, it is customary to introduce the density parameter
$\Omega_{gw}$ to characterize the strength of the gravitational
wave field \cite{Allen1997}, \cite{glpps2001},
\cite{Grishchuk2003}. $\Omega_{gw}$ is related to the power
spectrum $P_h(k)$ by the relation
\bea
\Omega_{gw}(k) = \frac{\pi^2}{3} \left( \frac{k}{k_H}\right)^2
P_h(k)\label{definitionofOmega}
\ena
where $k_H = 2\pi H_o/c$, and $H_o$ is the present day Hubble
constant. The quantity $\Omega_{gw}$ is the present day ratio of
energy density of gravitational waves (per unit logarithmic
interval in $k$) to the critical density of the Universe
$\rho_{crit} = 3c^2H_o^2/8\pi G$.

For simplicity, below we shall assume a simple power law behaviour
for density parameter $\Omega_{gw} = \Omega_{gw} (k_o) \cdot
(k/k_o)^{n_T}$. Although restricted, this form of spectrum is a
good approximation for a large variety of models in gravitational
wave frequency range of our interest. For example, this type of a
power spectrum arises due to the evolution of relic gravitational
waves with a (primordial) spectral index equal to $n_T$,
(i.e.~$P_h(k)|_{prim}\propto k^{n_T}$). The flat, scale invariant
power spectrum (also known as Harrison-Zeldovich power spectrum)
corresponds to $n_T=0$. In general the power law spectrum for
$\Omega_{gw}$ just assumes the absence of features in the spectrum
of gravitational waves at the wavelengths of our interest.

Let us consider electromagnetic radiation from a distant quasar.
Expression (\ref{deltaalphasquare}) allows us to calculate the
expected angular fluctuation in the position of this quasar caused
by a stochastic background of gravitational waves. In order to
proceed we require to specify the limits of integration $k_{min}$
and $k_{max}$ in (\ref{deltaalphasquare}). $k_{min}$ and $k_{max}$
determine the frequency range of gravitational waves that can be
probed by precision interferometry. The lower limit $k_{min}$ is
determined by the time duration of observations $T_{obs}$,
$k_{min} \approx 2\pi/cT_{obs}$. The upper limit $k_{max}\approx
2\pi/c\delta t$ is determined by the time resolution of the
observations $\delta t$, and we shall assume $\delta t\ll T_{obs}$
(i.e.~$k_{max}\gg k_{min}$). Let $D$ be the distance to the
quasar, which we shall assume is comparable to the Hubble length,
i.e.~$D \approx L_H = cH_o^{-1}$. We shall be working under the
assumption $kD = 2\pi D/\lambda_{gw} \gg 1$, corresponding to the
reasonable condition that the gravitational waves of our interest
have a wavelengths much shorter than $L_H$.

As can be seen from expression (\ref{deltaalphasquare4}) the
behaviour of the transfer function $\Delta\tilde{\alpha}^2(k)$
depends on value of the quantity $5\pi\epsilon^3kL_H$. In order to
analyze the various possibilities let us introduce
\bea
\epsilon_* = \left(5\pi k_{min}L_H\right)^{-1/3} \approx 2.3 \cdot
10^{-4},
\ena
where we have assumed $k_{min} = 2\pi/cT_{obs}$, and
$T_{obs}=10~yrs$. In the above expression, and elsewhere below we
set $H_o = 75~\frac{km}{sec}/Mpc$ for numerical evaluations.

For the angular displacement $<\Delta\alpha^2>$, in the case
$\epsilon \ll \epsilon_{*}$, substituting
(\ref{deltaalphasquare4cosmological} into
(\ref{deltaalphasquare}), taking into account the definition
(\ref{definitionofOmega}) and integrating in the limits from
$k_{min}$ to $k_{max}$ we get
\bea
<\Delta\alpha^2> &=&
\frac{3}{40\pi^2}~\frac{\Omega_{gw}\left(k_o\right)}{(1-n_T/2)}\left[
\left(\frac{k_H}{k_{min}}\right)^2\left(\frac{k_{min}}{k_o}\right)^{n_T}
-
\left(\frac{k_H}{k_{max}}\right)^2\left(\frac{k_{max}}{k_o}\right)^{n_T}
\right]\left(\frac{(1+z)^2+1}{2}\right)
\nonumber
\\ &\approx&
\frac{3}{40\pi^2}~\frac{\Omega_{gw}\left(k_{min}\right)}{(1-n_T/2)}\left(
\frac{T_{obs}}{T_{H}} \right)^2\left(\frac{(1+z)^2+1}{2}\right),
~~~~~\textrm{for}~\epsilon\ll\epsilon_{*}.
\label{alphaofomega_smallepsilon}
\ena
In the opposite case of $\epsilon\gg\epsilon_*$ we get
\bea
<\Delta\alpha^2> &=&
\frac{3\Omega_{gw}\left(k_o\right)\epsilon^3}{(1-n_T)}\left[
\left(\frac{k_H}{k_{min}}\right)\left(\frac{k_{min}}{k_o}\right)^{n_T}
-
\left(\frac{k_H}{k_{max}}\right)\left(\frac{k_{max}}{k_o}\right)^{n_T}
\right]\left(1+z-{\sqrt{1+z}}\right)
\nonumber
\\ &\approx&
\frac{3\Omega_{gw}\left(k_{min}\right)\epsilon^3}{(1-n_T)}\left(
\frac{T_{obs}}{T_{H}} \right)\left(1+z-{\sqrt{1+z}}\right),
~~~~~\textrm{for}~\epsilon\gg\epsilon_{*}.
\label{alphaofomega}
\ena
In expressions (\ref{alphaofomega_smallepsilon}) and
(\ref{alphaofomega}) $T_H = H_o^{-1}$ is the Hubble time and $k_H
= 2\pi/L_H$. We have assumed $k_{max} \gg k_{min}$ and used
$k_H/k_{min} \simeq T_{obs}/T_H$. In the above expressions we
restrict our analysis to the case $n_T<1$ which covers most of the
practically interesting cases, including $n_T=0$ corresponding to
a flat primordial spectrum of relic gravitational waves. In
further evaluations below we shall set the redshift $z=4$, which
corresponds to a redshift with significant amount of quasar
sources available for observations.

The measurement of $<\Delta\alpha^2>$ for distant quasars by the
planned interferometric projects \cite{RadioAstron},
\cite{Interferometers} would be able to constrain either
$\Omega_{gw}$ or $\Omega_{gw}\epsilon^3$, depending on the value
of $\epsilon$ compared with $\epsilon_*$. A null result in the
measurement of $<\Delta\alpha^2>$, in the case $\epsilon\ll
\epsilon_*$, would place the following limit on the energy density
of gravitational waves $\Omega_{gw}$ (using expression
(\ref{alphaofomega_smallepsilon}))
\bea
\Omega_{gw} ~ \leq ~ 4.0\cdot10^{-4}\left[ \left( 1 - n_T/2
\right)\left( \frac{\Delta\alpha_{rms}}{1~ \mu\textrm{as}}
\right)^2 \left( \frac{10~{\textrm{yrs}}}{T_{obs}} \right)^2
\right].
\label{omegalimit}
\ena
The above expression serves as the (weakest) upper limit on
$\Omega_{gw}$ that can be set by precision interferometry
measurements, irrespective of the value $\epsilon$ (since, as will
become clearer from the expression below, for values of
$\epsilon\gg\epsilon_*$ this upper limit only becomes more
stringent). In the case $\epsilon\gg \epsilon_*$, from
(\ref{alphaofomega}), we get the following upper limit for the
quantity $\Omega_{gw}\epsilon^3$
\bea
\Omega_{gw}\epsilon^3 ~ \leq ~ 3.7\cdot10^{-15}\left[ \left( 1 -
n_T \right)\left( \frac{\Delta\alpha_{rms}}{1~ \mu\textrm{as}}
\right)^2 \left( \frac{10~{\textrm{yrs}}}{T_{obs}} \right)
\right].
\label{omegaepsilon3limit}
\ena
In expressions (\ref{omegalimit}) and (\ref{omegaepsilon3limit})
we have introduced $\Delta\alpha_{rms} = \sqrt{<\Delta\alpha^2>}$,
which is the root mean square of the angular resolution of the
interferometer. This precision is around $1~\mu as$ for the
currently planned interferometers, and reaches $0.4~\mu as$ for
the proposed MAXIM x-ray interferometer. It is worth noting that,
at this angular resolution, the upper limit on $\Omega_{gw}
\lesssim 6.4\cdot10^{-5}$ that can be achieved by precision
interferometric measurements is comparable to the current limits
set by LIGO \cite{LIGO2005}, \cite{LIGO2006}.

\begin{figure}
\begin{center}
\includegraphics[width=14cm]{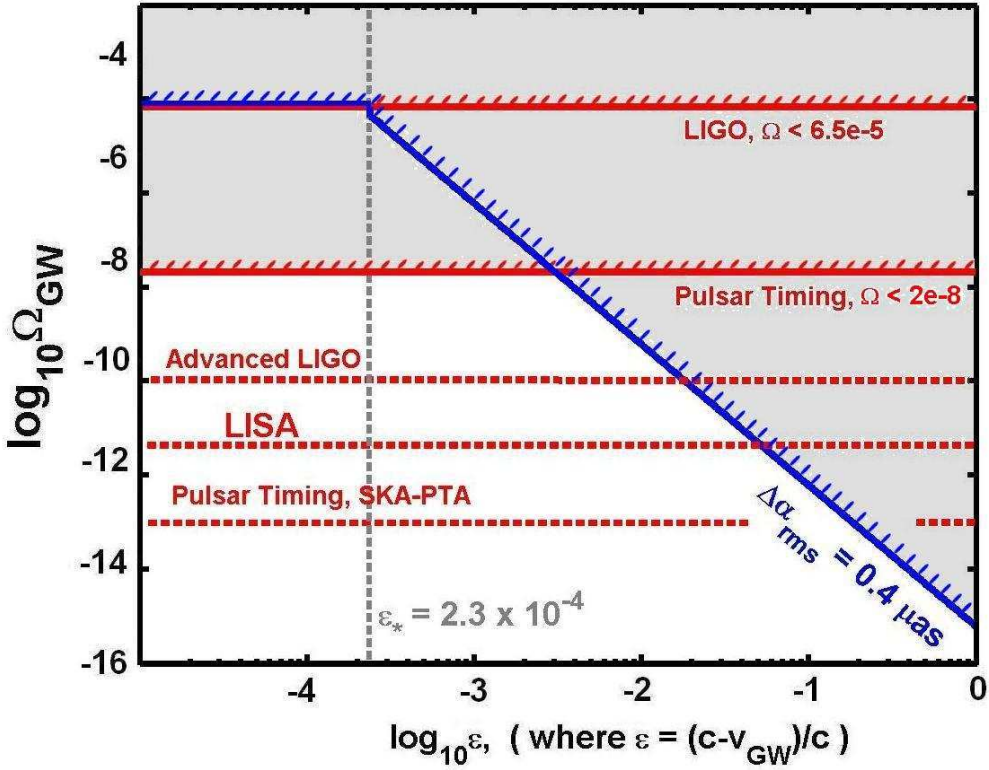}
\end{center}
\caption{The upper limit on the energy density $\Omega_{gw}$ and
velocity parameter $\epsilon$ achievable by an interferometric
observation of a source at redshift $z=4$ with resolution
$\Delta\alpha_{rms} = 0.4 ~\mu as$ and $T_{obs}=10~yrs$. The
shaded area shows the region that can be ruled out by precision
interferometry together with other existing constraints. The
horizontal lines on the graph show various constraints on the
density parameter $\Omega_{gw}$ (solid hairy lines - current upper
bounds, dashed line - future sensitivity levels).}\label{figure2}
\end{figure}

Figure \ref{figure2} shows the constraints on energy density
parameter $\Omega_{gw}$ and the velocity parameter $\epsilon$
achievable with an angular resolution of $\Delta\alpha_{rms} =
0.4~\mu as$ promised by the MAXIM project \cite{Interferometers}.
The figure also shows the current constraints on the $\Omega_{gw}$
parameter \cite{LIGO2005}, \cite{LIGO2006}, \cite{PulsarTiming},
\cite{PulsarTiming3}, along with sensitivity levels of some of the
planned experiments \cite{AdvancedLIGO}, \cite{LISA},
\cite{PulsarTiming2}.

An independent measurement of $\Omega_{gw}$ (at a level below
$\Omega_{gw}\lesssim 10^{-5}$) by ground based interferometers
\cite{glpps2001}, planned space borne interferometer LISA
\cite{LISA}, or Cosmic Microwave Background anisotropy and
polarization measurements \cite{Keating2006a} would allow to place
direct constraints on velocity parameter $\epsilon$. In this case,
using (\ref{alphaofomega}), we can calculate the upper limit on
$\epsilon$ achievable by precision interferometric measurements
\bea
\epsilon ~ \leq ~ 3.3\cdot 10^{-2} ~ \left[ \left( 1 - n_T \right)
\left( \frac{10^{-10}}{\Omega_{gw}} \right) \left(
\frac{\Delta\alpha_{rms}}{1~ \mu\textrm{as}} \right)^2 \left(
\frac{10~{\textrm{yrs}}}{T_{obs}} \right) \right]^{\frac{1}{3}}.
\label{epsilonlimit}
\ena
This constraint corresponds to the region $\epsilon>\epsilon_*$ on
Figure \ref{figure2}.

In table \ref{table:models} we summarize the predictions for
$\Omega_{gw}$, and the corresponding upper limits on $\epsilon$,
for some of the viable models that generate a considerable amount
of stochastic gravitational wave backgrounds \cite{gangui2001},
\cite{hogan2006}, \cite{dgp2006}.

\begin{table}[ht]

\centering 
\begin{tabular}{| l | c | c | c |} 
\hline\hline 
Theoretical Model & Predicted $\Omega_{gw}$ & Upper Limit on $\epsilon$ & Upper Limit on $m_g$ \\ [0.5ex] 
\hline 
Relic gravitational waves, $n_T=0$ & $\Omega_{gw} \approx
10^{-14}$ & $\epsilon \lesssim 0.4$ &  $m_g \lesssim
5.2\cdot10^{-23}$ eV  \\
Relic gravitational waves, $n_T=0.2$ at $\nu=1$ Hz & $\Omega_{gw} \approx 10^{-10}$ & $\epsilon \lesssim 1.8\cdot10^{-2}$  &  $m_g \lesssim 2.3\cdot10^{-24}$ eV  \\
Local Strings & $\Omega_{gw} \approx 10^{-8}$ & $\epsilon \lesssim 3.9\cdot 10^{-3}$ &   $m_g \lesssim 5.1\cdot10^{-25}$ eV \\ 
Global strings & $\Omega_{gw} \approx 10^{-12}$ & $\epsilon \lesssim 8.4\cdot10^{-2}$ &  $m_g \lesssim 1.1\cdot10^{-23}$ eV  \\
Extended Inflation & $\Omega_{gw} \approx 10^{-9}$ & $\epsilon \lesssim 8.4\cdot10^{-3}$ &   $m_g \lesssim 1.1\cdot10^{-24}$ eV \\
$1^{st}$ order EW transitions & $\Omega_{gw} \approx 10^{-9}$ &
$\epsilon \lesssim 8.4\cdot10^{-2}$ &   $m_g \lesssim
1.1\cdot10^{-24}$ eV  \\ [1ex]
\hline 
\end{tabular}
\caption{An upper limit on the velocity parameter $\epsilon$, for
some viable models that predict a considerable gravitational wave
background, that can be placed by interferometric observations
with resolution $\Delta\alpha_{rms} = 0.4 ~\mu as$, and
observation time $T_{obs}=10~yrs$. The last column shows the upper
limits on the mass of the graviton $m_g$ (see equation
(\ref{gravitonmass})) for a fiducial wavelength
$\lambda_{gw}\approx 1~ly$.}
\label{table:models} 
\end{table}

\section{\label{sec:conclusion} Conclusion}

Although gravitational waves are yet to be detected, there are
currently upper limits placed by observations. The CMB places the
most stringent limits on gravitational waves of cosmological
origin of $\Omega_{gw} \leq 10^{-14}$ extrapolated to laboratory
scale frequencies. It is crucial that this limit depends on
extrapolation of data measured at the very long wavelengths
(comparable to the Hubble length) down to wavelengths of the order
of one light year and much lesser. Further more, these
extrapolations assume a spectral index close to zero, i.e.~a flat
(scale invariant) spectrum of primordial gravitational waves. At
frequencies relevant for precision interferometry observations,
the most stringent constraint is placed by the Pulsar Timing
measurements $\Omega_{gw} < 2 \times 10^{-8}$ at frequencies
$10^{-9} - {10}^{-7}$ Hz \cite{PulsarTiming3}, and LIGO results
$\Omega_{gw} < 6.5 \times 10^{-5}$ at frequencies 51-150 Hz
\cite{LIGO2005}, \cite{LIGO2006}. As can be seen from Table
\ref{table:models}, there are a host of viable theoretical models
that predict gravitational wave backgrounds above the sensitivity
levels of planned experiments like Advanced LIGO, LISA and
SKA-PTA. If these experiments detect a gravitational wave
background, the precision interferometry observations would be
able to place strong constraints on $\epsilon \lesssim 10^{-2}$.
Let us finally note that, these interferometric measurements would
be directly sensitive to gravitational waves with frequencies
$\nu_{gw}\lesssim2\pi/T_{obs}\approx 2\cdot10^{-8}$ Hz, which is
the frequency region that would be probed by LISA and SKA-PTA.

Along side the candidates for a statistically isotropic
gravitational wave background, we can also expect a significant
stochastic background of gravitational waves from galactic white
dwarf binaries \cite{CutlerThorne2001} which are expected to be a
dominant contribution to ``noise" in LISA. Consideration of
anisotropic gravitational wave backgrounds requires an approach
slightly different from considerations of Section
\ref{sec:arbitrarywavefield}, but it is reasonable to assume that
these sources could also place considerable upper limits on
$\epsilon$.

\section*{Acknowledgements}

The authors thank  B.~G.~Keating, W.~Zhao, L.~P.~Grishchuk and
M.~V.~Sazhin for discussions and useful suggestions.




\appendix

\section{Evaluation of the transfer function
\label{SurfingIntegrals}}

Let us evaluate the integral in expression
(\ref{deltaalphasquare3}) in the physically interesting case when
$\epsilon\rightarrow 0$ and $kD\gg1$. In this case the expression
for the transfer function can be separated into two distinctive
contributions
\bea
\Delta\tilde{\alpha}^2(k) = \Delta\tilde{\alpha}_{NR}^2(k) +
\Delta\tilde{\alpha}_{R}^2(k),
\ena
where $\Delta\tilde{\alpha}_{NR}^2(k)$ is the non-resonance
contribution
\bea
\Delta\tilde{\alpha}_{NR}^2(k) & = &
\frac{1}{16}\int\limits_{-1}^{1-\epsilon-\Delta\mu} d\mu ~ \left(
1-\mu^2 \right)^3 \left[\frac{\sin^2{\left\{ \frac{\pi
D}{\lambda_{gw}} \left(1 - \epsilon - \mu\right)
\right\}}}{\left(1 - \epsilon - \mu\right)^2}\right] \nonumber \\
& & + \frac{1}{16}\int\limits_{1-\epsilon+\Delta\mu}^{+1} d\mu ~
\left( 1-\mu^2 \right)^3 \left[\frac{\sin^2{\left\{ \frac{\pi
D}{\lambda_{gw}} \left(1 - \epsilon - \mu\right)
\right\}}}{\left(1 - \epsilon - \mu\right)^2}\right],
\label{NRintegral}
\ena
and $\Delta\tilde{\alpha}_{R}^2(k)$ is the resonance contribution
\bea
\Delta\tilde{\alpha}_{R}^2(k) & = &
\frac{1}{16}\int\limits_{1-\epsilon-\Delta\mu}^{1-\epsilon+\Delta\mu}
d\mu ~ \left( 1-\mu^2 \right)^3 \left[\frac{\sin^2{\left\{
\frac{\pi D}{\lambda_{gw}} \left(1 - \epsilon - \mu\right)
\right\}}}{\left(1 - \epsilon - \mu\right)^2}\right].
\label{Rintegral}
\ena
The quantity $\Delta\mu$ occurring in the limits of integration in
the above expressions is fixed by the condition for the resonance
to occur. This condition corresponds to the region, around
$\mu=1-\epsilon$, where the sine function undergoes a few
oscillations. Thus $\Delta\mu = N\lambda_{gw}/D = 2\pi N/kD$,
where $N$ is the number of oscillations of the sine function,
around the point $\mu=1-\epsilon$, included in evaluation of the
resonance. The value of $N$ is limited by the condition $\Delta\mu
= 2\pi N/kD\ll\epsilon$, implying $N\ll\epsilon kD/2\pi$. Since in
all our considerations we assume $\epsilon\ll 1$, and
$\epsilon^3kD\gg1$, the condition imposed on $N$ is consistent
with an additional condition $N\gg1$ that we shall assume.

When evaluating (\ref{NRintegral}), in the case of
$\epsilon\rightarrow 0$, we can neglect the second integral in
comparison with the first. In evaluation fo the remaining integral
we set $\epsilon = 0$. Thus, we get
\bea
\Delta\tilde{\alpha}_{NR}^2(k) & \approx &
\frac{1}{16}\int\limits_{-1}^{1} d\mu ~ \left( 1-\mu \right)\left(
1+\mu \right)^3 \sin^2{\left( \frac{k D}{2} \left(1 - \mu\right)
\right)} \nonumber \\ & =& \frac{1}{32}\int\limits_{-1}^{1} d\mu ~
\left( 1-\mu \right)\left( 1+\mu \right)^3
\left(1-\frac{}{}\cos{\left( {k D} \left(1 -
\mu\right)\right)}\right)\nonumber \\ & \approx &
\frac{1}{32}\int\limits_{-1}^{1} d\mu ~ \left( 1-\mu \right)\left(
1+\mu \right)^3 = \frac{1}{20},\label{NRintegral2}
\ena
where, assuming $kD\gg1$, we have explicitly separate out the
rapid oscillatory part and neglect it.

In order to evaluate (\ref{Rintegral}), in the case of
$\epsilon\rightarrow 0$ and $kD\gg1$, it is helpful to notice that
the factor $\left( 1-\mu^2 \right)^3$ in the right side of
(\ref{Rintegral}) is a slowly varying function over the range of
integration. Taking this factor (evaluated at $\mu=1-\epsilon$)
outside the integral we get the following approximation for the
resonance part of the transfer function
\bea
\Delta\tilde{\alpha}_{R}^2(k) & \approx &
\frac{1}{2}\epsilon^3\int\limits_{1-\epsilon-\Delta\mu}^{1-\epsilon+\Delta\mu}
d\mu ~ \left[\frac{\sin^2{\left\{ \frac{k D}{2} \left(1 - \epsilon
- \mu\right) \right\}}}{\left(1 - \epsilon - \mu\right)^2}\right]
 =
\frac{1}{4}\epsilon^3kD\int\limits_{-N\pi}^{+N\pi}dx\frac{\sin^2{x}}{x^2}\nonumber \\
~&\approx& ~
\frac{1}{4}\pi\epsilon^3kD\left(1-O\left(\frac{1}{N^2}\right)\right)
~ \approx ~ \frac{1}{4}\pi\epsilon^3kD.
\label{Rintegral2}
\ena

Finally, the total transfer function, given by the sum of the
non-resonance (\ref{NRintegral2}) and resonance parts
(\ref{Rintegral2}), has the following form
\bea
\Delta\tilde{\alpha}^2(k) \approx \frac{1}{20}\left[1 +\frac{}{}
5\pi\epsilon^3kD\right]. \label{AppendixAintegralTotal}
\ena

\section{The surfing effect in the presence of cosmological
evolution
\label{CosmologyInclusion}}

In Section \ref{sec:singlewave} when evaluating expression
(\ref{firstordersolution1}) for simplicity and clarity we had
neglected cosmological evolution of the gravitational wave
amplitude, i.e.~we set $a(\eta)=1$. When we incorporate the
cosmological expansion, it no longer becomes possible to evaluate
the integral in expression (\ref{firstordersolution1}) in terms of
elementary functions. Although this complicates the calculational
aspects of the problem, much of the physical aspects remain the
same as discussed in Sections \ref{sec:singlewave} and
\ref{sec:arbitrarywavefield}. Introducing
\bea
\gamma \equiv k(1-\epsilon-\tilde{k}_{i}e^i),
\label{definitionofgamma}
\ena
the expression (\ref{firstordersolution1}) (in the case of a
matter dominated universe governed by the scale factor
(\ref{mattermetric})) can be rewritten in the following form
\bea
\psi_1(\eta,x^i) = \frac{\omega_E  h_o}{2c} ~
e^ie^kp_{ik}~\gamma\eta_o^2\int\limits_{\gamma(\eta_o-D)}^{\gamma\eta_o}dx\frac{e^{ix}}{x^2},
\ena
The expression (\ref{deltaalpha1}) for the angular displacement
due to a gravitational wave modifies correspondingly to
\bea
\Delta\alpha = -\frac{i}{2}h_o ~
e^ie^kp_{ik}~\tilde{k}_rl^r~\gamma
k\eta_o^2\int\limits_{\gamma(\eta_o-D)}^{\gamma\eta_o}dx\frac{e^{ix}}{x^2}.
\label{deltaalphacosmology1}
\ena

At this point it is convenient to consider the non-resonance and
resonance contributions separately. Let us consider each in turn,
beginning with the non-resonance contribution. For this case,
$\gamma\eta_o \gtrsim \gamma(\eta_o-D) \gg 1$, and the integral in
expression (\ref{deltaalphacosmology1}) can be evaluated
asymptotically to give
\bea
\Delta\alpha_{NR} &\approx& -\frac{1}{2}h_o ~
e^ie^kp_{ik}~\tilde{k}_rl^r~\gamma
k\eta_o^2~\left.\left(\frac{e^{ix}}{x^2} +
O\left(\frac{1}{x^3}\right)\right)\right|_{\gamma(\eta_o-D)}^{\gamma\eta_o}.
\nonumber \\
& \approx & \frac{1}{2}h_o ~e^ie^kp_{ik}~\tilde{k}_rl^r~
e^{i\gamma
\eta_o}\left(\frac{\eta_o}{\eta_o-D}\right)^2\left[\frac{k}{\gamma}\left(e^{i\gamma
D}-\left(\frac{\eta_o-D}{\eta_o}\right)^2\right)\right]
\nonumber \\
& = & \frac{1}{2}h_o (1+z)~e^ie^kp_{ik}~\tilde{k}_rl^r~
e^{2i\gamma L_H}\left[\frac{e^{2i\gamma L_H
\left(1-(1+z)^{-1/2}\right)}-(1+z)^{-1}}{\left(1-\epsilon-\tilde{k}_ie^i\right)}\right],
\label{deltaalphacosmologyNR}
\ena
where we have used $\eta_o=2L_H$ and expression (\ref{Dofz}) to
relate $D$ and redshift $z$. The key difference between
(\ref{deltaalphacosmologyNR}) and (\ref{deltaalpha1}) is the
presence of the cosmological redshift factor $(1+z)$. This is the
reflection of the fact that the non-resonance part of
$\Delta\alpha$ probes the gravitational wave field at emission
when the field was a factor $(1+z)$ stronger than at present. The
non-resonance contribution to the transfer function
(\ref{deltaalphasquare2}) can be evaluated in a fashion similar to
the evaluation of $\Delta\tilde{\alpha}_{NR}^2$ in Appendix
\ref{SurfingIntegrals}. The result is given by
\bea
\Delta\tilde{\alpha}_{NR}^2(k) = \frac{1}{4\pi}\int d\Omega
\left.\frac{}{}\left| \Delta\tilde{\alpha}(k^i)
\right|^2\right|_{\epsilon=0} =
\frac{1}{20}\left[\frac{(1+z)^2+1}{2} \right].
\ena
Let us now turn to the resonance contribution in the transfer
function. This contribution can be estimated by setting
$\gamma(\eta_o-D)\lesssim \gamma\eta_o \ll 1 $. In this limit we
can approximate the expression (\ref{deltaalphacosmology1}) in the
following way
\bea
\left.\frac{}{}\Delta\alpha_{R}\right|_{\gamma\eta_o\rightarrow0}
&\approx& \frac{i}{2}h_o ~ e^ie^kp_{ik}~\tilde{k}_rl^r~\gamma
k\eta_o^2~\left.\frac{1}{x}\right|_{\gamma(\eta_o-D)}^{\gamma\eta_o}.
\nonumber \\
& = & -\frac{i}{2}h_o ~kD \sqrt{1+z}~e^ie^kp_{ik}~\tilde{k}_rl^r.
\label{deltaalphacosmologyR1}
\ena
Thus, in comparison with expression (\ref{deltaalpha2}) taken in
the limit $\gamma D\rightarrow0$, the expression
(\ref{deltaalphacosmologyR1}) has an extra factor $\sqrt{1+z}$.
Using this approximation we can estimate the resonance part of the
transfer function as follows
\bea
\Delta\tilde{\alpha}_{R}^2(k) & \approx &
\frac{1}{16}(1+z)\int\limits_{1-\epsilon-\Delta\mu}^{1-\epsilon+\Delta\mu}
d\mu ~ \left( 1-\mu^2 \right)^3 \left[\frac{\sin^2{\left\{
\frac{\pi D}{\lambda_{gw}} \left(1 - \epsilon - \mu\right)
\right\}}}{\left(1 - \epsilon - \mu\right)^2}\right]\nonumber \\
~&\approx & ~ \frac{1}{4}(1+z)\pi\epsilon^3kD \nonumber \\ &=&
\frac{1}{4}\pi\epsilon^3kL_H(1+z)\left(2-\frac{2}{\sqrt{1+z}}\right),
\label{deltaalphacosmologyR2}
\ena
where the resonance integral has been evaluated in the same manner
as in Appendix \ref{SurfingIntegrals}. Figure \ref{figureAppendix}
shows the comparison of the approximate integrand used in
(\ref{deltaalphacosmologyR2}) and the exact integrand calculated
numerically. As can be seen from the figure, expression
(\ref{deltaalphacosmologyR2}) gives a good approximation to the
exact result.

\begin{figure}
\begin{center}
\includegraphics[width=8cm]{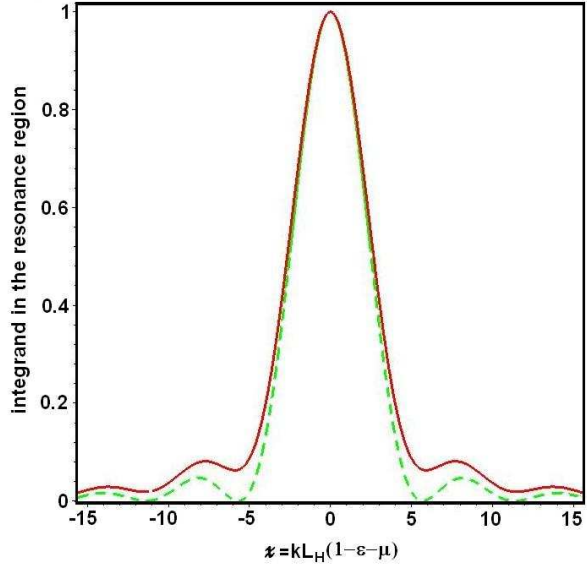}
\end{center}
\caption{The integrand figuring in the evaluation of the transfer
function $\Delta\tilde{\alpha}_{R}^2(k)$ for the case of
resonance, i.e.~$\mu\approx1-\epsilon$. The dashed line shows the
approximation used in (\ref{deltaalphacosmologyR2}), while the
solid line shows the integrand in the exact evaluation. The value
of redshift is set to $z=4$. The overall (common to both the
curves) normalization has been chosen
arbitrarily.}\label{figureAppendix}
\end{figure}

Combining expressions (\ref{deltaalphacosmologyNR}) and
(\ref{deltaalphacosmologyR2}) we get the expression for the total
transfer function
\bea
\Delta\tilde{\alpha}^2(k) &\approx&  \frac{1}{20}\left[
\left(\frac{(1+z)^2+1}{2}\right) +
5\pi\left(1+z\right)\epsilon^3kD\right] \nonumber \\
&=& \frac{1}{20}\left[ \left(\frac{(1+z)^2+1}{2}\right) +
5\pi\epsilon^3kL_H\left(1+z\right)\left(2-\frac{2}{\sqrt{1+z}}\right)\right].
\ena


\end{document}